\documentclass[sigconf]{acmart}

\AtBeginDocument{%
  }

\usepackage{listings}
\usepackage{xcolor}

\begin{document}

\title{LookSync: Large-Scale Visual Product Search System for AI-Generated Fashion Looks}

\author{Pradeep M}
\email{pradeep.m@glance.com}
\affiliation{%
  \institution{Glance}
  \city{Bangalore}
  \country{India}
}

\author{Ritesh Pallod}
\email{ritesh.pallod@glance.com}
\affiliation{%
  \institution{Glance}
  \city{Bangalore}
  \country{India}
}

\author{Satyen Abrol}
\email{satyen.abrol@glance.com}
\affiliation{%
  \institution{Glance}
  \city{Bangalore}
  \country{India}
}

\author{Muthu Raman T}
\email{muthu.raman@glance.com}
\affiliation{%
  \institution{Glance}
  \city{Bangalore}
  \country{India}
}

\author{Ian Anderson}
\email{ian.anderson@glance.com}
\affiliation{%
  \institution{Glance}
  \city{London}
  \country{United Kingdom}
}

\renewcommand{\shortauthors}{Pradeep et al.}

\begin{abstract}
Generative AI is reshaping fashion by enabling virtual looks and avatars making it essential to find real products that best match AI-generated styles. We propose an end-to-end product search system that has been deployed in a real-world, internet scale which ensures that AI-generated looks presented to users are matched with the most visually and semantically similar products from the indexed vector space. The search pipeline is composed of four key components: query generation, vectorization, candidate retrieval, and reranking based on AI-generated looks. Recommendation quality is evaluated using human-judged accuracy scores. The system currently serves more than 350,000 AI Looks in production per day, covering diverse product categories across global markets of over 12 million products. In our experiments, we observed that across multiple annotators and categories, CLIP~\cite{radford2021learning} outperformed alternative models by a small relative margin of 3--7\% in mean opinion scores~\cite{huynh2011rating}. These improvements, though modest in absolute numbers, resulted in noticeably better user perception matches, establishing CLIP~\cite{radford2021learning}  as the most reliable backbone for production deployment.

\end{abstract}




\maketitle

\section{Introduction}
The rapid growth of e-commerce platforms and AI-driven personalization has increased the demand for scalable and accurate product search systems. Traditional e-commerce search pipelines are designed primarily for scenarios where users explicitly search for products or discover them through recommendations. However, modern consumers increasingly expect a more immersive and personalized experience such as visualizing how products would look on them before making a purchase.

An increasingly popular technique to address this gap is Virtual Try-On (VTON), which allows users to visualize themselves in specific garments at the point of discovery. Unlike conventional search pipelines that rely on structured catalog metadata or keyword matching, our use case requires matching products to AI-generated images that may not correspond to any real item in the catalog. This creates a unique challenge: retrieving the closest possible products in terms of visual style, category, and contextual fit.

To solve this, we have developed a large-scale, end-to-end product search system that indexes over 12 million products across diverse catalogs and geographies, generating high-dimensional embeddings for product images with continuously updated metadata. On the query side, the system processes AI-generated looks, extracts product attributes, and performs reverse mapping to retrieve the most visually and semantically similar items from the indexed vector space. In production, the system maintains an average end-to-end latency of under 1 second for online search requests while supporting over 350,000 AI-generated looks daily. This demonstrates the trade-off achieved between precision (human MOS consistently >3.5) and speed, ensuring the system remains accurate and responsive at internet scale.

\section{Background}
Visual search for product discovery within catalogs has seen growing adoption in recent times. For example, leading search platforms enables user to search various e-commerce catalogs when they upload an image. Also, Fashion retailers have enabled users to do visual search products from their catalog in fashion retail. However, these solutions are not tuned for AI-generated outfits that mix styles or create variations not present in inventory.

Recent development in vision–language models like CLIP~\cite{radford2021learning}, 
FashionCLIP~\cite{chia2022contrastive}, Fashion SigLIP~\cite{zhu2025generalized}, 
and DINOv2~\cite{xiao2024florence2} have significantly improved the ability to map image content into high-dimensional embedding spaces where semantic and visual similarities can be measured. These models enable content-based retrieval, supporting large-scale indexing in vector databases for  search. CLIP~\cite{radford2021learning} and SigLIP~\cite{zhu2025generalized} learn joint image–text representations, making them robust for multi-modal queries, while DINOv2~\cite{xiao2024florence2} offers strong self-supervised image feature extraction, making them strong contender for the search system we have. Although these architectures have demonstrated success in research and niche deployments, their application at scale in production environments with millions of products and frequent catalog updates presents operational and performance challenges.

One of the key challenges we face with AI-generated looks, is that these images may depict ensembles that do not exist as exact products in the catalog, necessitating approximate visual and semantic matching rather than direct lookup. This led us to develop a system that leverages deep visual embeddings, scalable vector search, and reranking to provide relevant alternatives in real time. By integrating these components into the application, we bridge the gap between synthetic outfit generation and tangible product discovery in global, multi-catalog retail settings.

\section{Glance AI And Product Matching}

Glance AI, generative AI shopping platform, aimed at transforming the e-commerce landscape by offering personalized, AI-powered shopping experiences directly on users' lock screens. Once users are onboarded to the app, they can upload their selfies and eventually try on their avatars on the reference images to get their AI looks which they can shop through the app.

\begin{figure}[htp]
    \centering
    \includegraphics[width=7cm]{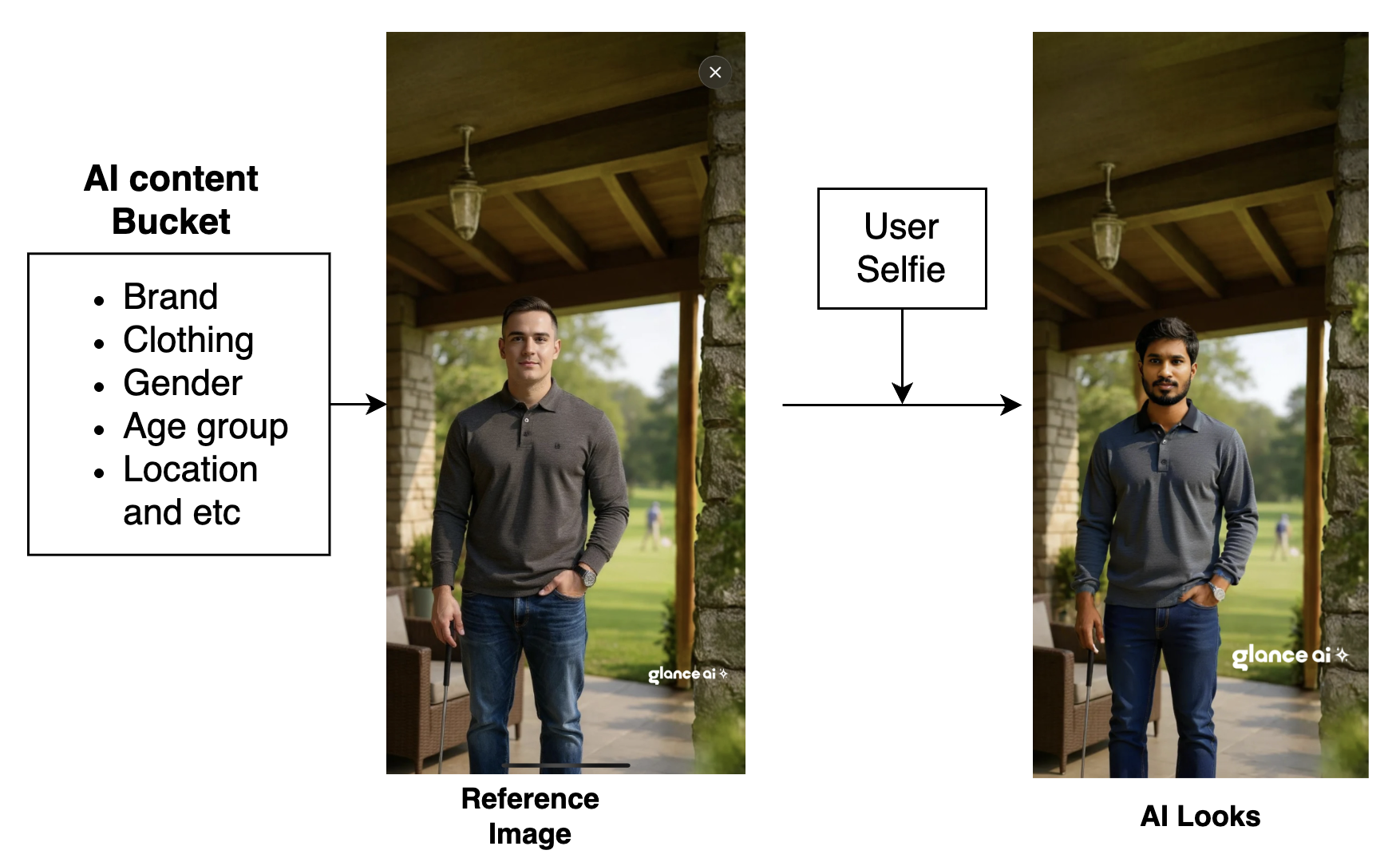}
    \caption{An image of a reference image to AI looks transition in the Glance AI app}
    \label{fig:ref_img_final1}
\end{figure}
The above diagram explains how the reference images are converted to the AI Looks based on the user selfie.  However, these generated images often contain clothing items or accessories that don’t exist exactly in the catalog either because the design is unique to the AI generation.

This is where product matching becomes essential. The goal is not only to find exact matches when available but also to approximate the AI look with the closest possible alternatives in terms of visual style, color, category, and fit. By doing this, it is ensured that every AI-generated look is shoppable, even when the exact product does not exist. The matching system bridges the gap between creative AI generation and the practical realities of retail inventory, maintaining a seamless shopping experience for the user.

A demo video of product matching in Glance AI available at \url{https://youtu.be/DZdlWmTUwjc}. 
The Glance AI app is publicly available on the app stores:
Android — \url{https://play.google.com/store/apps/details?id=com.glance.ai} \;
and iOS — \url{https://apps.apple.com/in/app/glance-ai-shop-with-ai/id6742974181}.

\section{Method}

\subsection{Ingestion of Product Catalogs}
Our system integrates with multiple vendors who supply product catalog information from a variety of retailers. We maintain a continuous ingestion pipeline that listens to a message queue, which receives product updates whenever a new product is added or existing metadata is modified.

When an update packet is received, the system checks whether it contains only metadata changes or new products. For metadata updates, we directly refresh the stored information. For new products, the associated images are converted into high-dimensional embeddings using the  CLIP~\cite{radford2021learning} (Contrastive Language–Image Pre-Training) model. These embeddings are stored in a vector database alongside the product metadata to enable efficient similarity search. Our systems have ingested more than 12 million products across multiple geographies.

\begin{figure}[htp]
    \centering
    \includegraphics[width=7cm]{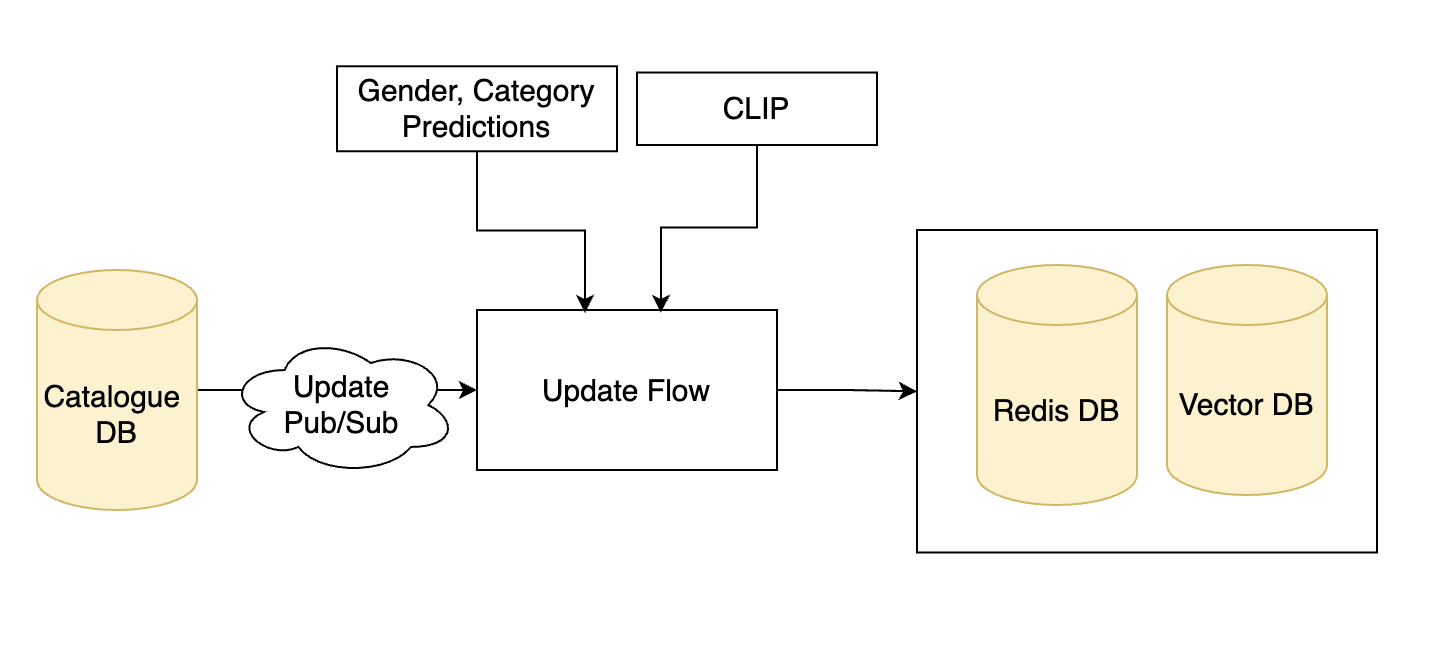}
    \caption{Ingestion of product catalogs into the product search system}
    \label{fig:update_flow_final1}
\end{figure}
\subsubsection{Product Metadata Enrichment}
During ingestion, the system enriches the metadata with standardized attributes such as category and gender. This standardization is necessary because the raw data arrives in different formats from multiple retailers. These enriched attributes improve downstream filtering, allowing the search pipeline to quickly narrow down to the most relevant candidates.

To further optimize performance, these standardized attributes are also stored in a Redis cache. This reduces the overhead of frequent vector database queries, especially for internal operations such as data migration, sanity checks, and coverage analysis.

\subsection{Product Search}
When the system receives an AI-generated look as input, it attempts to match it to the most visually and semantically similar products in the indexed vector space. This matching process is performed through a multi-stage pipeline consisting of query generation, vectorization, candidate retrieval, and reranking.

\begin{figure}[htp]
    \centering
    \includegraphics[width=8cm,height=6cm]{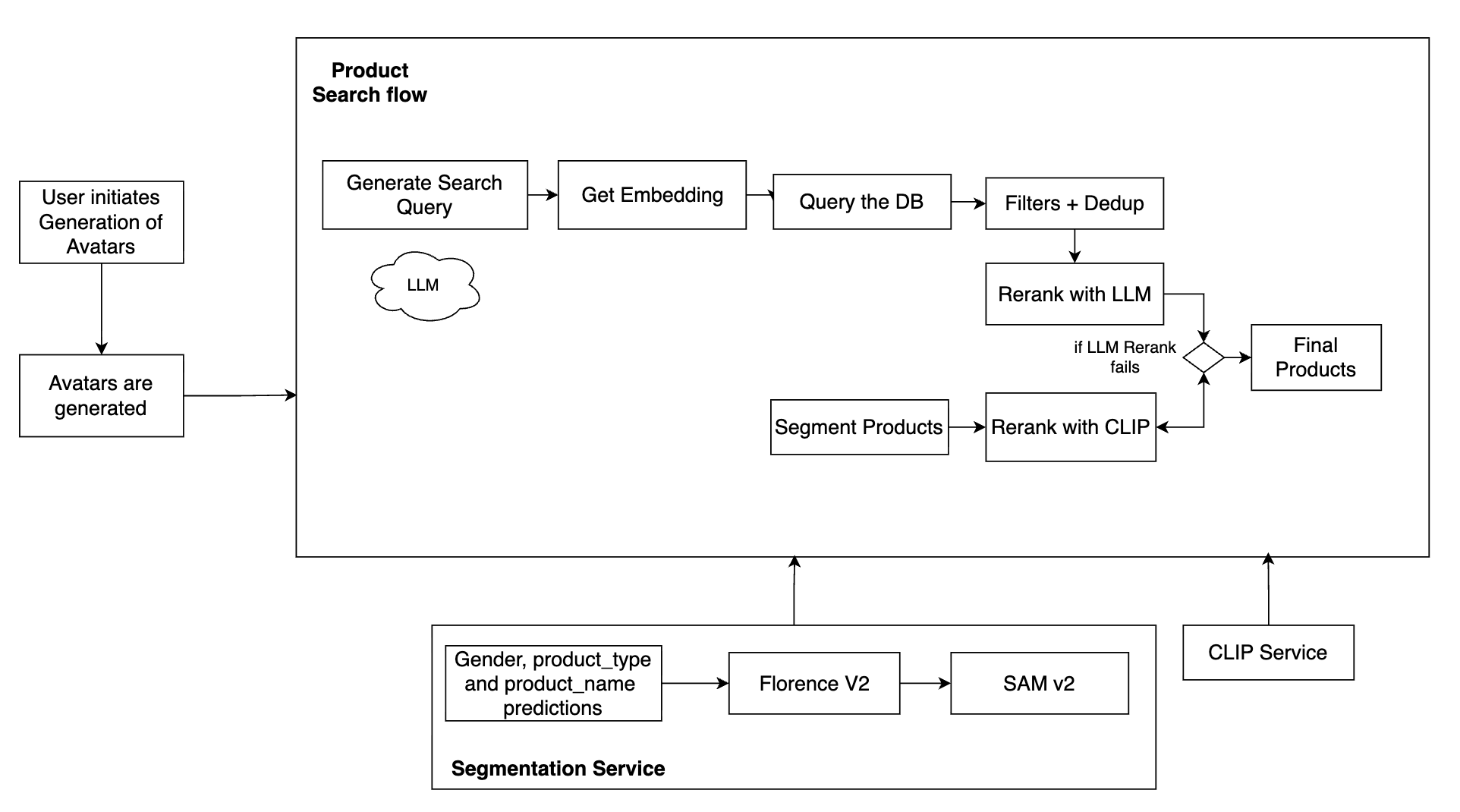}
    \caption{Architecture diagram of the product search system}
    \label{fig:arch_struct_final2}
\end{figure}

\subsubsection{Query Generation}
In this stage, the reference images extracted from AI-generated looks are passed to a large language model (LLM) to produce enhanced search queries that best describe the products being worn. The prompt design for the LLM is optimized to ensure the generated descriptions capture detailed product characteristics, enabling accurate text-based searches.

The output of this stage is a structured dictionary mapping each detected product layer to its corresponding descriptive query. For example:

\begin{lstlisting}
{
'outermost_topwear': "Men's charcoal grey polo shirt, solid, button-down collar, long sleeves, straight hem, casualwear.",
'bottomwear': "Men's indigo blue denim jeans, solid, straight leg, medium wash, five-pocket styling, casualwear.",
'accessory_1': "Men's silver wristwatch, round face, metal band, analog display, everyday wear."
}
\end{lstlisting}

These detailed, layer-specific descriptions serve as high-quality search queries for downstream embedding generation and retrieval.

\subsubsection{ CLIP~\cite{radford2021learning} (Contrastive Language–Image Pre-Training)}
Each generated query is then passed through the  CLIP~\cite{radford2021learning} model. 
In particular, we employ the ViT-H/14 variant trained on the LAION-2B dataset 
(CLIP-ViT-H-14-laion2B-s32B-b79K), as it offers strong performance on large-scale visual-language tasks. CLIP~\cite{radford2021learning} is a multimodal model trained to understand and align text with images in the same vector space. This means it can take a text description (like the search queries above) and produce an embedding that lives in the same space as our product image embeddings. Since our catalog images have already been embedded with  CLIP~\cite{radford2021learning} during ingestion, this allows us to directly compare the query embeddings to the product embeddings for similarity.

\subsubsection{Candidate Retrieval}
Using these embeddings, we query our vector database to fetch the closest matching products. Once candidates are retrieved, we deduplicate highly similar products and apply hard filters — such as brand, size, and price — if the user has explicit preferences or if we infer them from past interactions.

\subsubsection{Reranking}
We then pass the top-k candidates for each product group to an LLM, which reranks them based on fine-grained visual and semantic similarity. This step ensures the product most similar to the AI look consistently appears at the top.

\subsubsection{Fallback Reranker (Product Segmentation)}
If the LLM-based reranker fails, we fall back to a segmentation-based approach using Facebook’s SAM v2 and Microsoft’s Florence model. These segment the individual products directly from the AI-generated look. Each segment is embedded using  CLIP~\cite{radford2021learning}, and candidates are reranked purely based on cosine similarity.

\begin{figure}[htp]
    \centering
    \includegraphics[width=7cm]{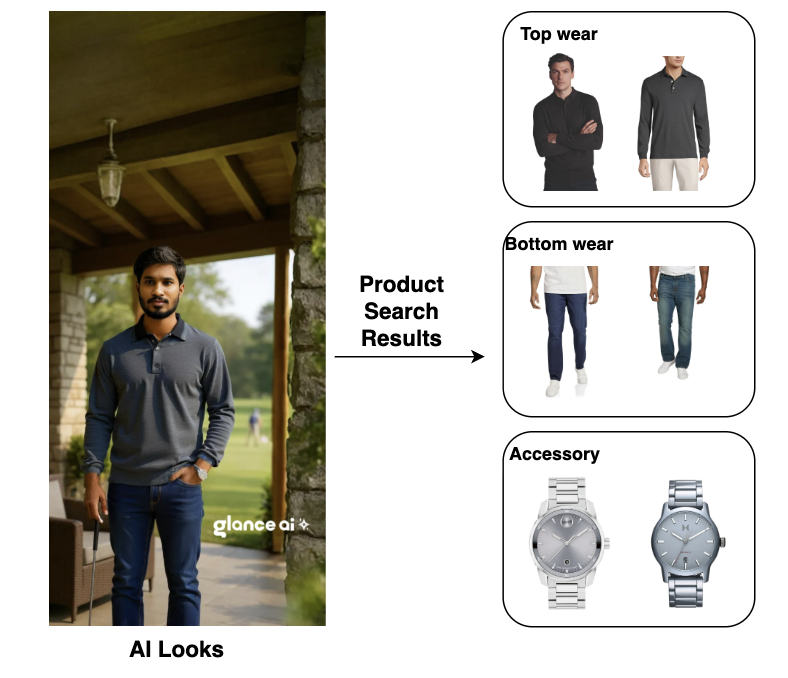}
    \caption{Outputs of product search for a AI generated look}
    \label{fig:output_resize_final}
\end{figure}
\section{Data and Experiments}
\subsection{Data}
Our current product search system works at a massive scale indexing roughly 12 million products from multiple retailers across geographies like India, the USA and Japan. These products cover a broad range of categories, from top wear and bottom wear to accessories and footwear. Every product image is converted into an embedding using CLIP~\cite{radford2021learning} (Contrastive Language–Image Pre-Training) and stored in a vector database. The system runs continuously, ingesting new products and updating existing ones in real time as vendors push changes. These 12 million products are tagged across 350,000 AI Looks across multiple geographies everyday.

\subsection{Experiments}
We’ve gone through several iterations to refine our product search accuracy. Over time, we tested multiple embedding models like  CLIP~\cite{radford2021learning}, FashionCLIP~\cite{chia2022contrastive}, Fashion SigLIP~\cite{zhu2025generalized}, and DINOv2~\cite{xiao2024florence2} each bringing its own strengths. Models like DINO v2 and Fashion-SigLIP showed great results in fine-grained aspects such as color and pattern detection. However,  CLIP~\cite{radford2021learning} stood out as the most consistent performer overall, balancing visual and semantic matching in a way that worked best for our needs.

\begin{table}[h!]
\centering
\setlength{\tabcolsep}{2pt}
\caption{Mean Opinion Scores (MOS)~\cite{huynh2011rating} across models (scale: 1–5). The highest score in each row is highlighted in bold.}
\label{tab:mos}
\small
\begin{tabular}{l l c c c}
\toprule
\textbf{Annotator} & \textbf{Gender} & \textbf{CLIP ViT-H/14} & \textbf{Fashion SigLip} & \textbf{DINO V2} \\
\midrule
1 & Male   & 2.70 & \textbf{2.80} & 2.50 \\
  & Female & \textbf{2.56} & 2.19 & 2.00 \\
\midrule
2 & Male   & \textbf{3.79} & 3.65 & 3.60 \\
  & Female & \textbf{4.12} & 4.03 & 3.88 \\
\midrule
3 & Male   & \textbf{3.55} & 3.50 & 3.52 \\
  & Female & 3.50 & \textbf{3.72} & 3.54 \\
\midrule
4 & Male   & \textbf{3.34} & 3.03 & 2.90 \\
  & Female & \textbf{2.79} & 2.76 & 2.80 \\
\midrule
5 & Male   & \textbf{3.90} & 3.70 & 3.65 \\
  & Female & \textbf{4.18} & 3.88 & 4.03 \\
\bottomrule
\end{tabular}
\end{table}

To measure accuracy, we relied on manual evaluations by human judges, using mean opinion scores (MOS)~\cite{huynh2011rating}. Evaluators considered multiple factors, including color match, fit, sleeve type, fabric type, pattern, and overall look. Table~\ref{tab:mos} summarizes key experiments. These continuous cycles of testing and feedback informed our production setup, which now strikes a solid balance between precision, coverage, and speed. From these results, we observe that CLIP~\cite{radford2021learning} consistently outperformed other models, making it the most reliable choice for deployment.

\section{User Interface}

The Product Search System was integrated with the Glance AI App. Users can view the products by onboarding to the App by uploading their selfie and some general information which are used to create their avatars. 

For each AI-generated look, a shop icon is displayed, allowing users to view visually and semantically similar products from the catalog. For reference images, the corresponding products are listed directly below the image.

Once the user clicks on the products, they are redirected to a product display page, where information about the products is presented, including price, stock, size chart and similar products. If user clicks the buy now they are redirected to the affiliate pages for purchase.

\begin{figure}[htp]
    \centering
    \includegraphics[width=6cm]{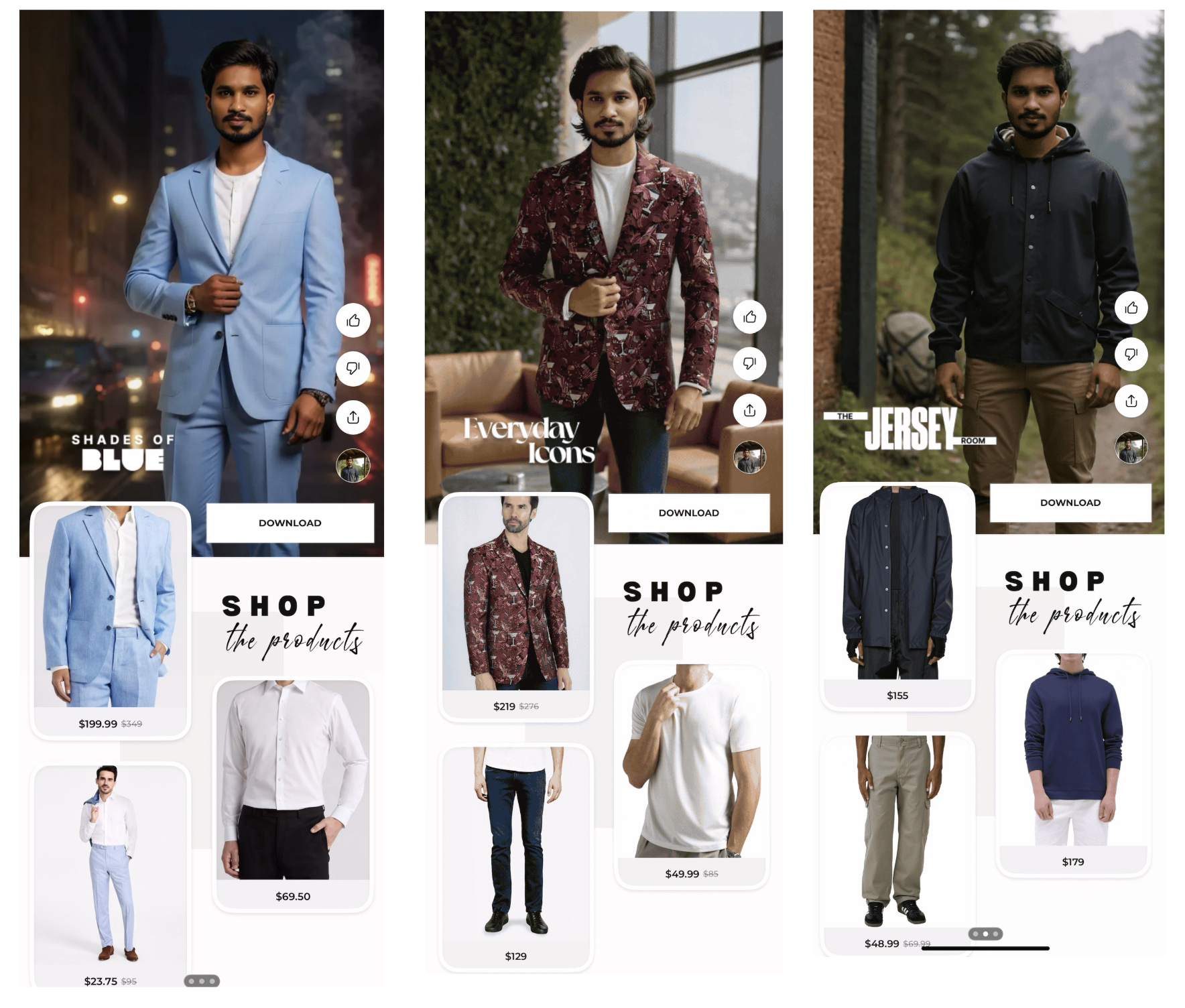}
    \caption{Multiple images of AI generated looks along with similar products}
    \label{fig:app_1}
\end{figure}

\section{Conclusion}

We presented a large-scale product search system that matches AI-generated looks to visually and semantically similar products. Our approach matches the AI generated looks to the similar products in our catalogue. It uses large language model (LLM) to generate accurate search queries, which are converted to  CLIP~\cite{radford2021learning} embeddings and these embeddings are in turn used to search the Vector DB to get the closest match, which are further reranked based on LLM to improve the product similarity to a higher extend.

Deployed in production, the system indexes over 12 million products and serves more than 350,000 AI Looks per day, achieving high human-judged mean opinion scores across diverse categories including top wear, bottom wear, accessories and footwear. Unlike traditional metadata-driven search, our method handles cases without exact catalog matches, allowing for approximate but relevant recommendations.

To serve real-time user interactions, we maintain both online and offline search pipelines. Offline jobs handle large-scale indexing, deduplication, and bulk updates, while online jobs power interactive queries from end-users. With aggressive caching and optimized search, the end-to-end system is responsive enough for immersive, AI-driven shopping experiences at internet scale.
\bibliographystyle{ACM-Reference-Format}
\bibliography{references}


\begin{thebibliography}{5}


\ifx \showCODEN    \undefined \def \showCODEN     #1{\unskip}     \fi
\ifx \showISBNx    \undefined \def \showISBNx     #1{\unskip}     \fi
\ifx \showISBNxiii \undefined \def \showISBNxiii  #1{\unskip}     \fi
\ifx \showISSN     \undefined \def \showISSN      #1{\unskip}     \fi
\ifx \showLCCN     \undefined \def \showLCCN      #1{\unskip}     \fi
\ifx \shownote     \undefined \def \shownote      #1{#1}          \fi
\ifx \showarticletitle \undefined \def \showarticletitle #1{#1}   \fi
\ifx \showURL      \undefined \def \showURL       {\relax}        \fi
\providecommand\bibfield[2]{#2}
\providecommand\bibinfo[2]{#2}
\providecommand\natexlab[1]{#1}
\providecommand\showeprint[2][]{arXiv:#2}

\bibitem[Chia et~al\mbox{.}(2022)]%
        {chia2022contrastive}
\bibfield{author}{\bibinfo{person}{Patrick~John Chia}, \bibinfo{person}{Giuseppe Attanasio}, \bibinfo{person}{Federico Bianchi}, \bibinfo{person}{Silvia Terragni}, \bibinfo{person}{Ana~Rita Magalh{\~a}es}, \bibinfo{person}{Diogo Goncalves}, \bibinfo{person}{Ciro Greco}, {and} \bibinfo{person}{Jacopo Tagliabue}.} \bibinfo{year}{2022}\natexlab{}.
\newblock \showarticletitle{Contrastive language and vision learning of general fashion concepts}.
\newblock \bibinfo{journal}{\emph{Scientific Reports}} \bibinfo{volume}{12}, \bibinfo{number}{1} (\bibinfo{year}{2022}), \bibinfo{pages}{18958}.
\newblock
\href{https://doi.org/10.1038/s41598-022-23052-9}{doi:\nolinkurl{10.1038/s41598-022-23052-9}}


\bibitem[Huynh-Thu et~al\mbox{.}(2011)]%
        {huynh2011rating}
\bibfield{author}{\bibinfo{person}{Quan Huynh-Thu}, \bibinfo{person}{Marie-Neige Garcia}, \bibinfo{person}{Filippo Speranza}, \bibinfo{person}{Philip Corriveau}, {and} \bibinfo{person}{Alexander Raake}.} \bibinfo{year}{2011}\natexlab{}.
\newblock \showarticletitle{Study of Rating Scales for Subjective Quality Assessment of High-Definition Video}.
\newblock \bibinfo{journal}{\emph{IEEE Transactions on Broadcasting}} \bibinfo{volume}{57}, \bibinfo{number}{1} (\bibinfo{date}{March} \bibinfo{year}{2011}), \bibinfo{pages}{1--14}.
\newblock
\showISSN{1557-9611}
\href{https://doi.org/10.1109/TBC.2010.2086750}{doi:\nolinkurl{10.1109/TBC.2010.2086750}}


\bibitem[Radford et~al\mbox{.}(2021)]%
        {radford2021learning}
\bibfield{author}{\bibinfo{person}{Alec Radford}, \bibinfo{person}{Jong~Wook Kim}, \bibinfo{person}{Chris Hallacy}, \bibinfo{person}{Aditya Ramesh}, \bibinfo{person}{Gabriel Goh}, \bibinfo{person}{Sandhini Agarwal}, \bibinfo{person}{Girish Sastry}, \bibinfo{person}{Amanda Askell}, \bibinfo{person}{Pamela Mishkin}, \bibinfo{person}{Jack Clark}, \bibinfo{person}{Gretchen Krueger}, {and} \bibinfo{person}{Ilya Sutskever}.} \bibinfo{year}{2021}\natexlab{}.
\newblock \showarticletitle{Learning Transferable Visual Models From Natural Language Supervision}.
\newblock \bibinfo{journal}{\emph{arXiv preprint arXiv:2103.00020}} (\bibinfo{year}{2021}).
\newblock
\href{https://doi.org/10.48550/arXiv.2103.00020}{doi:\nolinkurl{10.48550/arXiv.2103.00020}}


\bibitem[Xiao et~al\mbox{.}(2024)]%
        {xiao2024florence2}
\bibfield{author}{\bibinfo{person}{Bin Xiao}, \bibinfo{person}{Haiping Wu}, \bibinfo{person}{Weijian Xu}, \bibinfo{person}{Xiyang Dai}, \bibinfo{person}{Houdong Hu}, \bibinfo{person}{Yumao Lu}, \bibinfo{person}{Michael Zeng}, \bibinfo{person}{Ce Liu}, {and} \bibinfo{person}{Lu Yuan}.} \bibinfo{year}{2024}\natexlab{}.
\newblock \showarticletitle{Florence-2: Advancing a Unified Representation for a Variety of Vision Tasks}. In \bibinfo{booktitle}{\emph{Proceedings of the IEEE/CVF Conference on Computer Vision and Pattern Recognition (CVPR)}}.
\newblock
\urldef\tempurl%
\url{https://openaccess.thecvf.com/content/CVPR2024/papers/Xiao_Florence-2_Advancing_a_Unified_Representation_for_a_Variety_of_Vision_CVPR_2024_paper.pdf}
\showURL{%
\tempurl}


\bibitem[Zhu et~al\mbox{.}(2025)]%
        {zhu2025generalized}
\bibfield{author}{\bibinfo{person}{Rongtao Zhu}, \bibinfo{person}{Han Wang}, \bibinfo{person}{Mohan~Kumar Jayaraman}, \bibinfo{person}{Junting Pan}, \bibinfo{person}{Yogesh Gowda}, \bibinfo{person}{Pablo Badilla}, \bibinfo{person}{Minsu Cho}, \bibinfo{person}{Kaicheng Yu}, \bibinfo{person}{Rui Luo}, \bibinfo{person}{David Chan}, \bibinfo{person}{Alexander Kirillov}, \bibinfo{person}{Piotr Doll{\'a}r}, {and} \bibinfo{person}{Christoph Feichtenhofer}.} \bibinfo{year}{2025}\natexlab{}.
\newblock \bibinfo{title}{Generalized Contrastive Learning with Flexible Margins}.
\newblock
\showeprint[arxiv]{2404.08535}~[cs.CV]
\href{https://doi.org/10.48550/arXiv.2404.08535}{doi:\nolinkurl{10.48550/arXiv.2404.08535}}
\newblock
\shownote{arXiv:2404.08535}.


\end{thebibliography}
\end{document}